\documentstyle[epsf,fullpage,12pt]{article}

\newcommand{\q}{k}
\newcommand{\proof}{ {\bf Proof:} }
\newcommand{\level}{\mbox{$\theta$}}
\newcommand{\cL}{\mbox{$\cal L$}}
\newcommand{\curlyF}{\mbox{$\cal F$}}
\newcommand{\curlyI}{\mbox{$\cal I$}}

\newcommand{\supp}{\mbox{supp}}

\newtheorem{theorem}{Theorem}[section]
\newtheorem{definition}[theorem]{Definition}
\newtheorem{lemma}[theorem]{Lemma}
\newtheorem{corollary}[theorem]{Corollary}

\date{\underline{In press}; Submitted Nov. 12, 1996}

\title{Multiscale Computation with Interpolating Wavelets}

\author{ 
Ross A. Lippert\thanks{ Department of Mathematics Room 2-342,
Massachusetts Institute of Technology, Cambridge, MA 02139,
{\protect\verb+ripper@mit.edu+}, supported by a grant from
the Liberty Mutual Group Fund for scholarships and financial aid
for graduate mathematics students.  },
T.A. Arias\thanks{ Department of Physics Room 12-110,
Massachusetts Institute of Technology, Cambridge, MA 02139,
{\protect\verb+muchomas@mit.edu+}, supported in part by the MRSEC Program
of the National Science Foundation under award number DMR 94-00334,
the Alfred P. Sloan Foundation under award number BR-3456.}, 
and
Alan Edelman\thanks{ Department of Mathematics Room 2-380,
Massachusetts Institute of Technology, Cambridge, MA 02139,
{\protect\verb+edelman@math.mit.edu+}, supported by a fellowship from
NSF Grant 9404326-CCR.  }
}

\begin{document}

\countdef\pageno=0 \pageno=0

\maketitle

\bibstyle{plain}

Classifications: 65D05, 65D30, 82A71, 81C06, 81C07


\vspace{2.5 in}

\thanks{Computational
support provided by the MIT Xolas project and general support provided
by the State Street Bank Science Partnership
Fund at MIT}

\newpage

\begin{center}
\vspace{1.0in}

{\Large Multiscale Computation with Interpolating Wavelets}

\vspace{3.0 in}

Ross A. Lippert \\
905 Main St. \#10 \\
Cambridge, MA \\
O2139 \\
ripper@mit.edu \\
(617) 661-2879 \\

\end{center}

\newpage

\begin{abstract}

Multiresolution analyses based upon interpolets, interpolating scaling
functions introduced by Deslauriers and Dubuc, are particularly
well-suited to physical applications because they allow {\em exact}
recovery of the multiresolution representation of a function from its
sample values on a {\em finite} set of points in space.  We present a
detailed study of the application of wavelet concepts to physical
problems expressed in such bases.  The manuscript describes algorithms
for the associated transforms which, for properly constructed grids of
variable resolution, compute correctly without having to introduce
extra grid points.    We demonstrate that for the application of
local homogeneous operators in such bases, the non-standard multiply
of Beylkin, Coifman and Rokhlin\cite{BeylkinCoifmanRokhlin:91} also
proceeds exactly for inhomogeneous grids of appropriate form.  To
obtain less stringent conditions on the grids, we generalize the
non-standard multiply so that communication may proceed between
non-adjacent levels.  The manuscript concludes with timing comparisons
against na\"{\i}ve algorithms and an illustration of the
scale-independence of the convergence rate of the conjugate gradient
solution of Poisson's equation using a simple preconditioning.

\end{abstract}

\section{Introduction and Motivation}

Wavelets offer a means of approximating functions that allows
selective refinement.  If regions of an image or a signal have
exceptionally large variations, one need only store a set of
coefficients, determined by function values in neighborhoods of those
regions, in order to reproduce these variations accurately.  In this
way, one can have approximations of functions in terms of a basis that
has spatially varying resolution.  This approach reduces the memory
storage required to represent functions and may be used for data
compression.

Physical applications often involve multiple physical fields which
interact in space with non-linear couplings.  Efficient
implementations must minimize not only storage to represent the these
fields but also the processing required to describe their
interactions.  It is highly desirable to perform the needed operations
with a fixed, limited number of floating point operations for each
expansion coefficient and to minimize the number of points in space at
which physical interactions must be evaluated.

As a concrete example of a realistic application, consider the
computation of the quantum mechanical electronic structure of a
collection of atoms in three dimensions.  For other examples of
physical applications, the reader may wish to consult
\cite{BertoluzzaNaldi:96}, \cite{BeylkinKeiser:95},
\cite{FrohlichSchneider:94}, {\em et al.}.  Arias and coworkers
\cite{ChoAriasJoannopoulosLam:93},\cite{AriasChoLamTeter.proceedings:95},
and other works which have appeared after the original submission of
this manuscript nearly one year ago \cite{WeiChou},\cite{TymczakWang},
have studied the use of multiresolution bases in quantum mechanical
computations.  (For a review, see \cite{Arias.unpublished:97}.)  It is
a consequence of quantum physics that, near atomic nuclei, electronic
wave functions vary rapidly and, that far from the nuclei, the wave
functions tend to be much more smooth.  From this observation, one can
anticipate that the fine scale coefficients in a multiresolution
analysis of the electronic wave functions and associated physical
fields will be significant only near the atomic cores, allowing for
truncation.  This problem is thus an ideal candidate for
multiresolution techniques.

Within density functional theory \cite{HohenbergKohn:64}, the quantum
physics of electrons and nuclei involves two types of fields, the
Schr\"odinger wave function $\left\{\psi_i(r)\right\}$ for each
electron $i$ and the electrostatic potential $\phi(r)$ arising from
the average electronic density $n(r)\equiv \sum_i |\psi_i(r)|^2$.
Within the local density approximation (LDA) \cite{KohnSham}, the
solution for the correct values of these fields is obtained at the
saddle-point of lowest energy of the Lagrangian functional
\begin{eqnarray} \label{eqn:saddle}
\cL_{LDA}(\{\psi_i\},\phi) & = & \frac{1}{2} \sum_i f {\int{d^3r\, ||\nabla
\psi_i(r) ||^2}} + \int{d^3r \, V_{\mbox{nuc}}(r) n(r)} +
\int{d^3r\,\,\epsilon_{xc}(n(r)) n(r)} \cr & - & \int{d^3r\,\phi(r) n(r)} -
\frac{1}{8\pi} \int{d^3r\, ||\nabla \phi(r)||^2}.
\end{eqnarray}
Here, we work in units $\hbar=m=e=1$, $V_{\mbox{nuc}}(r)$ is the total
potential which each electron feels due to the presence of the atomic
nuclei, and $\epsilon_{xc}(n)$ is a non-algebraic function known only
through tabulated values.  (For a review of density functional theory,
see \cite{PayneTeterAllanAriasJoannopoulos:92}.)

In practice, one finds the fields $\left\{\psi_i(r)\right\}$,
$\phi(r)$ by
\begin{itemize}
\item expanding the fields in terms of unknown coefficients
within some basis set
\begin{eqnarray}
\psi_i(x) & = & \sum_\alpha c_{\alpha,i} b_\alpha(x) \label{eqn:exps} \\
\phi(x) & = & \sum_\alpha d_\alpha b_\alpha(x); \nonumber
\end{eqnarray}
\item evaluating Eq. (\ref{eqn:saddle}) in terms of the unknown
coefficients $c$ and $d$;
\item determining analytically the gradients
of the resulting $\cL_{LDA}(c,d)$ with respect to those coefficients;
and
\item proceeding with conjugate gradients to locate the saddle point.
\end{itemize}
All follows directly once one has expressed the Lagrangian as a
function of the expansion coefficients.

In doing this, we note that each term represents a local coupling in
space, but that one coupling, $\phi(r) n(r)$, is cubic in the field
coefficients $c$ and $d$,
and another, $\epsilon_{xc}(n(r)) n(r)$, is only known in terms of
tabulated values.  Expanding the product of two wavelets in terms of
wavelets on finer levels would make possible the treatment of the cubic
coupling to some level of approximation.  (See, for example,
\cite{BeylkinKeiser:97}.)  However, this route becomes
increasingly difficult for higher order interactions and is hopeless
for non-algebraic or tabulated interactions, such as
$\epsilon_{xc}(n(r))$.  For higher order interactions it is natural,
and for non-algebraic and tabulated interactions necessary, to
evaluate the interactions at some set of points in space and then
recover expansion coefficients for the result.  One then relies upon
the basis set to provide interpolation for the behavior at non-sample
points.

The benefits of both truncated wavelet bases and interpolation on
dyadically refined grids are given by the use of interpolating scaling
functions \cite{Dubuc:86}, \cite{BeylkinSaito:92},
\cite{BeylkinSaito:93}, \cite{DeslauriersDubuc:89},
\cite{Donoho.unpublished:92} (or interpolets
\cite{AriasYesil.unpublished:96}, \cite{Arias.unpublished:97}), which
are functions with the following properties (from
\cite{Donoho.unpublished:92}, pp. 6-7).

Let $\phi(x)$ be an interpolet, then
\begin{list}{}
\item (INT1) {\bf cardinality:} $\phi(k) = \delta_{0,k}$ for all $k \in Z^n$
\item (INT2) {\bf two-scale relation:} $\phi(x/2) = \sum_{y\in Z^n} c_y \phi(x-y)$
\item (INT3) {\bf polynomial span:} For some integer $m \ge 0$, any polynomial 
$p(x)$ of degree $m$ can be represented as a formal sum
$\sum_{y\in Z^n} a(y) \phi(x-y)$.
\end{list}

Cardinality allows the fast conversion between uniform samples and
interpolating scaling functions and has subtle yet profound
consequences for the resulting multiresolution basis.  In particular,
as is evident from our algorithms below, the expansion coefficient for
a basis function on a particular scale is independent of the samples
of the function for points associated with finer scales.
Consequently, the expansion coefficients which we obtain for functions
maintained in our basis are identical to what would be obtained were
function samples available on a complete grid of {\em arbitrarily}
fine resolution.  This eliminates all error in the evaluation of
non-linear, non-algebraic and tabulated interactions beyond the
expansion of the result in terms of a finite set of basis functions.

The $c_y$ in the two-scale relation are referred to as scaling
coefficients, and cardinality actually implies that $c_y = \phi(y/2)$.
The two-scale relation allows the resolution to vary locally in a
mathematically consistent manner.

The polynomial span condition captures, in some sense, the accuracy of
the approximation.  By cardinality, we actually have $a(y) = p(y)$.
We shall call $m$ the polynomial order.

Interpolets thus can be thought of as a bridge between computations with
samples on dyadically refined grids and computations in a
multiresolution analysis.  The former point of view is useful for
performing local nonlinear operations, while the latter is useful for
the application of local linear operators.

This manuscript explores $O(N)$ algorithms that calculate transforms
and linear operators for grids of variable resolution but return, for
the coefficients considered, {\em exactly} the same results as would
be obtained using a full, uniform grid at the highest resolution {\em
without} the need to introduce artificial temporary augmentation
points to the grid during processing.  We thus show that with
relatively mild conditions on the variability of the resolution
provided by the grid, interpolet bases provided the ultimate economy
in the introduction of grid points: only as many samples in space need
be considered as functions used in the basis.  The four transforms
(forward, inverse, and the dual to each) mapping between coefficients
and functions samples which we discuss here are particular to
interpolet bases.  For the application of operators in such bases, we
show that the familiar non-standard multiply of Beylkin, Coifman and
Rokhlin\cite{BeylkinCoifmanRokhlin:91} shares with the transforms the
property of correctness without the need to introduce additional grid
points.  Furthermore, we weaken the condition on grid variability by
using a modification of the non-standard multiply.  We generalize the
non-standard multiply so that communication may proceed between nearby
but non-adjacent levels and thereby obtain less stringent conditions
on the variability of the grid.  All of theoretical results in this
manuscript are presented in a general $d$-dimensional space.
Illustrative examples for the purpose of discussion will be given in
$d=1$ and $d=2$ dimensions.  The examples of applications in the final
section will be in $d=3$ dimensions.  Our focus is entirely on
interpolet bases, and so it remains an open question whether these
results hold true or can be adapted to more general wavelet systems.

Our organization is as follows.  In Sections 2 and 3, we explain how
to manipulate and construct interpolet expansions and some aspects of
how well they perform.  These sections will present nothing new to the
experienced wavelet user, but will explain our notational conventions
and recapitulates common theorems (\cite{Daubechies:CPAM-41-909},
\cite{SaitoBeylkin:93}, \cite{Daubechies:TLW-92},
\cite{Strang.book:96}, et al.)  for wavelet novices.  In Section 4, we
describe how nonuniform bases can be conceptualized in the framework
of interpolet expansions and
then use our results to develop algorithms for the transforms.
Section 5 details the algorithm for $\nabla^2$ and other operators.
Section 6 gives some practical details for the reader interested in
implementing these algorithms.  Finally, Section 7 compares, in three
dimensions, timings of these implementations with the timings of
na\"{\i}ve algorithms.  This final section also explores the
convergence of a preconditioned conjugate gradients algorithm in the
solution of Poisson's equation for the full three dimensional
electrostatic potential arising from the nuclei in the nitrogen
molecule.

\section{Introduction to Interpolets}

There is a unique set of interpolets on $R$ having symmetry and
minimal support for a given polynomial order $m=2l-1$ (the {\em
Deslauriers-Dubuc functions} \cite{DeslauriersDubuc:89}).  These are
the functions with which this article is primarily concerned (our
results carry over to more general interpolets, and no use will
actually be made of minimal support or symmetry).

To determine the $c_y$'s, one sets $c_{2j}=\delta_{m0}$ and $c_y =
c_{-y}$.  One may solve the Vandermonde system,
$$
\pmatrix{1 & 1 & \cdots & 1 \cr
	 1 & 3^2 & \cdots & (2l-1)^2 \cr
	 1 & 3^4 & \cdots & (2l-1)^4 \cr
	   &     & \cdots &  \cr
	 1 & 3^{2l-2} & \cdots & (2l-1)^{2l-2} \cr}
\pmatrix{c_1 \cr c_3 \cr . \cr . \cr c_{2l-1} \cr} = \pmatrix{\frac{1}{2} \cr 0 \cr . \cr . \cr 0 \cr}
$$
to obtain the remaining $c_y$'s.  These
coefficients satisfy the conditions for polynomial order $2l-1$.

The scaling coefficients for $m=1$ are
$$c_{-1} = c_1 = 0.5, c_0 = 1,$$
and for $m=3$ (the example used for Figure 1.) they are
$$c_{-3} = c_3 = -\frac{1}{16}, c_{-1} = c_1 = \frac{9}{16}, c_0 = 1.$$

One may then take tensor products of $\phi$'s and $c_y$'s to form
interpolets in higher dimensions.

\subsection{Interpolet Multiresolution Analysis}

We are concerned with recursive representations of
functions from samples at integer points on both uniform and
refined grids.  There are many definitions which make the
exposition more clear.

\begin{definition}{}
For $k >0$, let $C_k = 2^k Z^n$, and let $D_k = C_{k-1}-C_k$.  
For $k\le 0$, let $C_k = Z^n$, and $D_k = \emptyset$.
\end{definition}

We consider $C_k$ to be the set of coarse lattice points on the lattice 
$2^{k-1} Z^n$ and $D_k$, the detail lattice points, to be those points on 
$2^{k-1} Z^n$ which are not coarse.
Note:  $D_k \cup C_k = C_{k-1}$, and $Z^n = C_k \cup D_k \cup D_{k-1} \cup \cdots \cup D_1$ is a partition of $Z^n$.

\begin{definition}{}
We let $\level_k(y) = \mbox{min}(k,m)$ where $m$ is the largest
integer such that $2^m$ divides all of the components of $y$.
We call $\level_k(y)$ the {\em level} of the point $y$.
\end{definition}
Given the partition $Z^n = C_k \cup D_k \cup D_{k-1} \cup 
\cdots \cup D_1$, we have
$$\level_k(y) = \left\{ \begin{array}{lc}
k, & y \in C_k \\  l-1, & y \in D_l.
\end{array}\right.$$

\begin{definition}{}
Let $S \subset Z^n$.
Let $\phi(x)$ be an interpolet.
Let $\curlyI_\q(\phi,S)$ be
the space of functions given by formal sums of the form
 $\sum_{y \in S} a(y) \phi(\frac{x-y}{2^{\level_\q(y)}})$.
\end{definition}
Where $\phi$ and $S$ are understood, we may simply write $\curlyI_\q$.

\begin{definition}{}
Let $S \subset Z^n$.
Let $\curlyF_\q(S)$ be the vector space of $R$- or $C$- valued functions on $Z^n$ which are zero at any point not in $S$ (i.e. with support contained in $S$).
\end{definition}
Where $S$ is understood,
we may simply write $\curlyF_\q$.  Note: $\curlyF_\q(S) =
\curlyF_\q(S\cap C_k) \oplus \curlyF_\q(S\cap D_k) \oplus
\curlyF_\q(S\cap D_{k-1}) \oplus \cdots \oplus \curlyF_\q(S \cap D_1)$.

The meaning of the $\q$ subscript will be established by the next
definition, which will link vectors in $\curlyF_\q$ with functions in
$\curlyI_\q$.  It is for this reason that while the $\curlyF$'s are
technically identical, they are semantically different.  In practice,
the $\curlyF$'s are the actual data structures being stored on the
computer.

\begin{definition}{}
Let $\phi(x)$ be an interpolet.
Let $\iota^\phi_\q: S \rightarrow \curlyI_\q(\phi,S)$ be defined by
$$\iota^\phi_\q y = \phi(\frac{x-y}{2^{\level_\q(y)} } ).$$
This definition extends linearly to
the mapping $\iota^\phi_\q: \curlyF_\q(S) \rightarrow
\curlyI_\q(\phi,S)$ defined by:
$$\iota^\phi_\q v = \sum_{y \in S} v(y) (\iota^\phi_\q y)$$
i.e.

$$(\iota^\phi_\q v)(x) = \sum_{y\in S\cap C_\q} v(y)\phi((x-y)/2^\q) + \sum_{y\in S \cap D_{\q}} v(y) \phi((x-y)/2^{\q-1}) + \cdots + \sum_{y\in S \cap D_1} v(y) \phi(x-y).$$
\end{definition}

The set $S$ can be thought of as the set of points in a refined grid.
The $\iota^\phi_\q$ identifications allow one to think of $S$ 
as a set of functions, $\{(\iota^\phi_\q y) | y \in S \}$, which form
a basis of $\curlyI_\q(\phi,S)$.  We will sometimes refer to $S$
as a refined grid and sometimes as a basis with this identification
understood.

One should think of the $\curlyF_k$ as spaces of coefficients for
function expansions in the corresponding $\curlyI_k$ spaces, in the
basis $S$.  The $\iota^\phi_\q$ simply associate a set of coefficients in
$\curlyF_\q$ with a function in $\curlyI_\q$.  
When $\phi$ is understood, we may write just $\iota_\q$.

We are now in a position to state the basic theorems of interpolet
expansions on uniform grids.

\begin{theorem}{}
Let $\phi(x)$ be an interpolet on $R^n$.  Then 
each mapping $\iota_k:\curlyF_k(S) \rightarrow \curlyI_k(\phi,S)$
$(k=1,2,\dots)$ is an isomorphism.
\end{theorem}

\proof
Since the map $\iota_k$ is surjective, it is only necessary to show that
$\iota_k$ is injective.  By the definition, $\iota_k v = 0$ if and only if there
exist $v \in \curlyF_k$ such that
$$0 = \sum_{y\in S\cap C_k} v(y)\phi((x-y)/2^k) + \sum_{y\in S \cap D_{k}} v(y) \phi((x-y)/2^{k-1}) + \cdots + \sum_{y\in S \cap D_1} v(y) \phi(x-y).$$

Let $z\in S \cap C_k$.  By (INT1), we have $\phi((z-y)/2^k) =
\delta_{(z-y)/2^k, 0} = \delta_{z,y},$ for $y \in S\cap C_k$, and also
$\phi((z-y)/2^l) = 0,$ for $y \in S\cap D_l$.  So, $(\iota_k v)(z) =
v(z), z \in S\cap C_k$, therefore $v(z) = 0, z\in S\cap C_k$.  This
being so, one then has $(\iota_k v)(z) = v(z), z \in S\cap D_k$, so
$v(z) = 0, z\in S\cap D_k$.  Once again, $(\iota_k v)(z) = v(z), z \in
S\cap D_{k-1}$, thus we must have $v(z) = 0, z\in S\cap D_{k-1}$.
Continuing in this manner, we deduce that $v(y) = 0, y \in S$, thus
$v=0$.

$\Box$

\begin{corollary}{}
$$\curlyI_\q(S) = \curlyI_\q(S\cap C_k) \oplus \curlyI_\q(S\cap D_k) \oplus \cdots \oplus \curlyI_\q(S\cap D_1)$$
Since the sum is direct, the expansion is unique.
\end{corollary}
This corollary is a consequence of observations in the above proof.

\begin{theorem}{}
Let $\phi(x)$ be an interpolet on $R^n$.  Then 
$$\forall k, \curlyI_k(\phi,Z^n) = \curlyI_{k-1}(\phi,Z^n).$$
Consequently,
$$\forall k_1,k_2, \curlyI_{k_1}(\phi,Z^n) = \curlyI_{k_2}(\phi,Z^n).$$
\end{theorem}

\proof
To prove that $\curlyI_k \subset \curlyI_{k-1}$ we note that
$\curlyI_k \subset \curlyI_{k-1} \cup \curlyI_k(C_k)$.  Thus
we just need to show $\curlyI_k(C_k) \subset \curlyI_{k-1}$.

Translating by $z \in C_k$ and inserting powers of 2 where
appropriate, one can rewrite (INT2) for $\phi$ as
$$\phi((x-z)/2^k) = \phi((x-z)/2^{k-1}) + \sum_{y \in D_k} c_{y/2^{k-1}}
\phi((x-z-y)/2^{k-1}).$$
The terms in the right hand side are elements $\curlyI_{k-1}$.  Thus $\curlyI_k \subset \curlyI_{k-1}$.

To prove $\curlyI_{k-1} \subset \curlyI_k$ note that any element of
$\curlyI_{k-1}$ can be expressed as:
$$f(x) = \sum_{y \in C_{k-1}} a(y) \phi((x-y)/2^{k-1}) + \sum_{y \in D_{k-1}} a(y) \phi((x-y)/2^{k-2}) + \cdots \sum_{y \in D_1} a(y) \phi(x-y)\\$$
All the terms in this expansion but the first are elements of $\curlyI_k$.  
Since $C_{k-1} = C_k \cup D_k$ we may split the first sum up as,
$$ \sum_{y\in C_{k-1}}a(y) \phi((x-y)/2^{k-1}) = \sum_{y\in C_{k}}a(y) \phi((x-y)/2^{k-1})  + \sum_{y\in D_{k}}a(y) \phi((x-y)/2^{k-1})$$
The second term is also an element of $\curlyI_k$.

Rewriting (INT2) one has ($y \in C_k$):
$$\phi((x-y)/2^{k-1}) = \phi((x-y)/2^k) - \sum_{z \in D_k} c_{z/2^{k-1}} \phi((x-z-y)/2^{k-1}).$$
$y \in C_k, z \in D_k$, so $y+z \in D_k$, thus the right hand side is made up of
elements of $\curlyI_{k}$.  Thus, $\curlyI_{k-1} \subset \curlyI_k$.

$\Box$

\subsection{Interpolet Transforms}

\begin{corollary}{Interpolet Decomposition\\}
The set of isomorphisms $\iota_k$ induces a set of isomorphisms
$$J_{k_1,k_2} : \curlyF_{k_1} \rightarrow \curlyF_{k_2},$$
$J_{k_1,k_2} = \iota_{k_2}^{-1} \circ \iota_{k_1}$.
\end{corollary}
We refer to these isomorphisms as {\em interpolet transforms}.
It is our convention to let $J_k = J_{0, k}$ and $J_{-k} = J_{k, 0}$.
The reader will note that the $J_i$ are linear transformations on
the coefficient spaces, and are thus the primary object of 
computation.

We now turn to a study of the $J$'s.  It
is clear from the definition that for $k_1< k_2$, $J_{k_1, k_2} = 
J_{k_2-1, k_2} \circ J_{k_2-2, k_2-1} \circ \cdots \circ J_{k_1, k_1+1}$,
and similarly for $k_1 > k_2$.
Thus, we need only study the $J_{k, k+1}$ and $J_{k+1, k}$ mappings.

\begin{theorem}{Computation Theorem\\}
Let $v \in \curlyF_k(Z^n).$\\
$$(J_{k, k+1} v)(y) = \left\{ \begin{array}{lc}
v(y), & y \notin D_{k+1} \\  v'(y), & y \in D_{k+1}
\end{array}\right.$$
where $v'(y) = v(y) - \sum_{z \in C_{k+1}} c_{(y-z)/2^k} v(z).$

$$(J_{k+1, k} v)(y) = \left\{ \begin{array}{lc}
v(y), & y \notin D_{k+1} \\  v'(y), & y \in D_{k+1}
\end{array}\right.$$
where $v'(y) = v(y) + \sum_{z \in C_{k+1}} c_{(y+z)/2^k} v(z).$

\end{theorem} 
\proof
For $v \in \curlyF_k$ we have
$$\iota_k v = \sum_{y\in C_k} v(y)\phi((x-y)/2^k) + \sum_{y \in D_k} v(y) \phi((x-y)/2^{k-1}) + \cdots + \sum_{y \in D_1} v(y) \phi(x-y)$$
expanding the first term,
$$\iota_k v = \sum_{y\in C_{k+1}} v(y)\phi((x-y)/2^k) +\sum_{y\in D_{k+1}} v(y)\phi((x-y)/2^k) + \sum_{y \in D_k} v(y) \phi((x-y)/2^{k-1}) + \cdots $$
using (INT2), $\phi((x-z)/2^k) = \phi((x-z)/2^{k+1}) - \sum_{y \in D_{k+1}} c_{(y-z)/2^{k}} \phi((x-y)/2^k),$
$$\iota_k v = \sum_{y\in C_{k+1}} v(y)\phi((x-y)/2^{k+1}) +\sum_{y\in D_{k+1}} v'(y)\phi((x-y)/2^k) + \sum_{y \in D_k} v(y) \phi((x-y)/2^{k-1}) + \cdots$$
$$(\iota_{k+1}^{-1} \iota_k v)(y) = \left\{ \begin{array}{lc}
v(y), & y \notin D_{k+1} \\  v'(y), & y \in D_{k+1}
\end{array}\right.$$

where $v'(y) = v(y) - \sum_{z \in C_{k+1}} c_{(y-z)/2^k} v(z)$.

The proof for $J_{k+1, k}$ is similar.

$\Box$

Similar to what one might get with wavelets, we see that we can
compute the coefficients of interpolet expansions on uniform lattices
by a pyramid algorithm.  Computationally, this procedure can be
carried out by first computing the $D_1$ coefficients with $J_{0, 1}$,
then by computing the $D_2$ coefficients from the $C_1$ data with
$J_{1, 2}$, and so on.  In this sense, it is no different from
standard multiresolution decompositions.

A feature of the interpolet decomposition is that the
transformations all have a particular {\em lower triangular} form.  That
is, if we write $v \in \curlyF_k$ as a vector with its $C_{k+1}$
components first, its $D_{k+1}$ components second, and the rest of its
components third, then the transformation takes the form,
$$ J_{k, k+1} v = \pmatrix{I & 0 & 0 & 0 & 0 \cr M & I & 0 & 0 & 0 \cr
0 & 0 & I & 0 & 0 \cr 0 & 0 & 0 & I & 0 \cr 0 & 0 & 0 & 0 & I} 
\pmatrix{v_{C_{k+1}} \cr v_{D_{k+1}} \cr v_{D_k} \cr . \cr .}.$$
The inverse, $J_{k+1,k}$, is obtained by replacing $M$ with $-M$.

\section{Accuracy of Interpolet Approximation}
Given a function $f(x)$ on $R^n$, one can form an interpolet
approximation to $f$ by the formula: $$f(x) \sim \sum_{y \in C_0} f(y)
\phi(x-y) = \iota_0 f,$$ where $f$ on the right hand side is thought
of as a function restricted to $C_0 = Z^n$ 
(a more cumbersome but more precise notation 
is $\iota_0 \{f(y)\}|_{y\in Z^n}$).
  This approximation has
the property that $(\iota_0 f)(z) = f(z), z \in Z^n.$ 

Starting from the expansion $\iota_0 f \in \curlyI_0(\phi,Z^n)$ one
can construct equivalent expansions, $\iota_k (J_k f) \in
\curlyI_k(\phi,Z^n)$.  The coefficients $J_k f$ are referred to as
the interpolet transform of the function $f$.

If $f(x)$ is sufficiently smooth, then we can expect that the coefficients
$(J_k f)(y), y\in D_l, l\le k,$ will be small.  This statement is captured 
rigorously by the following lemma and theorem.  

\begin{lemma}{}
Let $\phi$ be an interpolet with polynomial order of $m$ then
$$p(x) \in \curlyI_N(\phi,C_N)$$
for any integer, $N$, and any polynomial, $p$, of degree $m$.
\end{lemma}
\proof
$p(2^N x)$ is a polynomial of degree $m$.
By (INT3), $p(x)$ can thus
be represented by a formal sum in $\curlyI_0(\phi,C_0)$, namely
$$p(2^N  x) = \sum_{y \in Z^n} p(2^N  y) \phi(x-y).$$
By changing variables, we may rewrite this as
\begin{eqnarray*}
p(x) &=& \sum_{y \in Z^n} p(2^N  y) \phi(x/2^N-y) \cr
&=& \sum_{y \in Z^n} p(2^N  y) \phi((x-2^N y)/2^N) \cr
&=& \sum_{y \in C_N} p(y) \phi((x-y)/2^N) \cr
&=& \sum_{y \in C_N} p(y) \phi((x-y)/2^{\level_N(y)}).\cr
\end{eqnarray*}

$\Box$

\begin{theorem}{(Residual Theorem)\\}
Let $f(x)$ be a polynomial function of degree $m$.
Let $\phi$ be an interpolet with a polynomial order of $m$.
Then 
$$(J_k f)(y) = \left\{ \begin{array}{lc}
f(y), & y \in C_k \\  0, & y \in D_l
\end{array}\right.$$
\end{theorem}
\proof

By (INT3), $f(x) \in \curlyI_0(\phi,C_0)$.  By
the lemma, we also have $f(x) \in \curlyI_k(\phi,C_k)$.  Recalling
that $J_k f$ gives the unique expansion coefficients of $f(x)$ in the
decomposition of $\curlyI_0(\phi,C_0)$ given by $\curlyI_k(\phi,C_k)
\oplus \curlyI_k(\phi,D_k) \oplus \cdots \oplus \curlyI_k(\phi,D_1),$
we see that the $\curlyI_k(\phi,D_l)$ coefficients must vanish, while
the $\curlyI_k(\phi,C_k)$ coefficients are given by the lemma, namely
$f(y)$ for $y\in C_k$.

$\Box$

The coefficients $(J_k f)(y), y \in D_l, l\le k,$ are called
{\em residuals}.  The Residual Theorem suggests that the magnitude of
the residual coefficients $(J_k f)$ at a point $y\in D_l$ are
expected to be proportional to the $(m+1)$th derivative of $f(x)$ at
$y.$ See Figure 2.

\section{Truncated Bases}

Typically, one truncates an expansion by eliminating elements of a
basis, setting their coefficients to $0$.  One then said to be working
in a truncated basis when one works within the subspace formed by the
remain basis elements.  In the notation of this paper, this
corresponds to taking expansions in $\curlyI_k(\phi,S)$ with
coefficients in $\curlyF_k(S)$.

One may also view a truncated basis as the set of expansions one gets
when the coefficients of some of the basis elements have been set to
``don't care'' values.  Mathematically, this is accomplished by
quotienting out the ``don't care'' elements.

\begin{definition}{}
Let
$$\curlyI_k^S = \curlyI_k(\phi,Z^n)/\curlyI_k(\phi,Z^n-S)$$
and
$$\curlyF_k^S = \curlyF_k(Z^n)/\curlyF_k(Z^n-S).$$
\end{definition}
When the identification of $F_k(S)$
with $F^S_k$ can be made will be dealt with later in this paper.  For
now, one may view these definitions as a trick to make the proofs less
complicated and for understanding exactly why and in what sense the
algorithms are correct.  Once again, we think of $\curlyF_k^S$ as a
grid on which the elements outside of $S$ have ``don't care'' values,
and $\curlyF_k(S)$ as a grid on which the elements outside of $S$ 
vanish.
The $\iota_k$ continue to be isomorphisms (since
$\iota_k(\curlyF_k(Z^n-S)) =
\curlyI_k(\phi,Z^n-S)$).

However, it is not necessarily true that $\curlyI_{k_1}^S = \curlyI_{k_2}^S$.
When this condition fails, then it is no longer possible to define
$J_{k_1, k_2} = \iota_{k_2}^{-1} \circ \iota_{k_1}$.

To be sure, one
could still define some sort of $J_{k_1, k_2}$ by setting 
the elements in $Z^n - S$ to zero, then applying the
full grid version of $J_{k_1, k_2}$, and then considering only
the elements in $S$ of the answer.
This definition by itself has some drawbacks.
Mathematically speaking, this is the same as
$J_{k_1, k_2} = p \circ \iota_{k_2}^{-1} \circ \iota_{k_1} \circ r$, where
$r:\curlyF_{k_1}^S \rightarrow \curlyF_{k_1}$, is some
lift to the full grid, and $p: \curlyI_{k_2}
\rightarrow \curlyI_{k_2}^S$ is the standard projection operator onto
the quotient.
Generally, if one follows this approach
one will no longer have $J_{k_1, k_2} = 
J_{k_2-1, k_2} \circ J_{k_2-2, k_2-1} \circ \cdots \circ J_{k_1, k_1+1}$
because $\curlyI_i^S$ are not all equal.
In terms of diagrams, where we once had
$$ \curlyF_{k_1} \stackrel{\iota_{k_1}}{\rightarrow} \curlyI_{k_1} = \curlyI_{k_2} \stackrel{\iota_{k_2}^{-1}}{\rightarrow} \curlyF_{k_2},$$
we now have
$$ \curlyF_{k_1}^S \stackrel{\iota_{k_1}}{\rightarrow} \curlyI_{k_1}^S \stackrel{?}{=} \curlyI_{k_2}^S \stackrel{\iota_{k_2}^{-1}}{\rightarrow} \curlyF_{k_2}^S.$$

What one needs is a condition on $S$ such that $\curlyI_{k_1}(Z^n-S)
=\curlyI_{k_2}(Z^n-S)$.  If this were true, then the definition of
the operator as $J_{k_1, k_2} = \iota_{k_2}^{-1} \circ \iota_{k_1}$ 
would actually be independent of the values of the
elements of $Z^n-S$.  In that case the quotient spaces are identical.

\begin{definition}{}
We say that the set $S$ is a {\em good basis} when it satisfies the
condition $\forall k_1, k_2, \curlyI_{k_1}(Z^n-S) = \curlyI_{k_2}(Z^n-S)$,
and thus $\curlyI^S_{k_1} = \curlyI^S_{k_2}.$
\end{definition}

To get a handle on this definition, one sees that
this is achieved when $\curlyI_{k}(Z^n-S) = \curlyI_{k+1}(Z^n-S)$.
For this to be so, whenever $y \in Z^n-S$, every $z$ such that
$\phi(x/2^{\level_N(z)}-z)$ 
is in the two-scale expansion for $\phi(x/2^{\level_N(y)}-y)$, 
must also be a member of $Z^n - S$. 

This can be captured in the following
table (in which we let $\level_k(y) =  \level_k(z)+1$).
\begin{center}
\begin{tabular}{l|cc}
in expansion? 		& $z \in S$ 	& $z \in Z^n-S$ \\ \hline
$y \in S$		& ok 	& ok \\ 
$y \in Z^n-S$		& not ok	& ok \\ 
\end{tabular}
\end{center}
The good basis condition for fast synthesis and reconstruction 
has also been discovered by Cohen and Danchin (see {\em S-trees} in
a coming work\cite{CohenDanchin:98}) which appeared after the
original submission of our manuscript.

For some of the algorithms presented, we may employ
additional conditions based on the supports of the functions
themselves (not just the support of their expansions).

\begin{definition}{}
We say that functions $f$ and $g$ {\em touch} whenever $\supp\{f\} \cap
\supp\{g\}$ has nonzero measure.

For $p > 0$ we say that $S$ has the {\em $p$-level touching} property
when it satisfies the condition, that
for $y \in Z^n-S$, $z \in Z^n,
\level_k(z) \le \level_k(y) - p,$ and $\iota_k y$ touches $\iota_k z$
implies $z \in Z^n - S.$
\end{definition}

A less formal way of phrasing this definition for the case of 1-level
touching is that if a level $l$ point,$y$, is a member of $Z^n - S$
then any point, $z$, at level $l-1$ or lower for which 
$\iota_k y$ touches $\iota_k z$ must also be in $Z^n-S$.
For 2-level touching, one only considers any
points at level $l-2$ or lower, and so on for $p$-level touching.

The allowed touching possibilities can be summarized the following 
table (in which we let $\level_k(y)
\ge \level_k(z)+p$).
\begin{center}
\begin{tabular}{l|cc}
touch? 		& $z \in S$ 	& $z \in Z^n-S$ \\ \hline
$y \in S$		& ok 	& ok \\ 
$y \in Z^n-S$		& not ok	& ok \\ 
\end{tabular}
\end{center}

One example of a 1-level touching good basis for one dimensional 3rd
order interpolets (the ones being used as an example) is the set of
interpolets centered at the points 
$$ S = C_n \cup \left( \bigcup^{n-1}_{l=0} \{-7\cdot 2^l,-5\cdot 2^l,\dots,5\cdot 2^l,7\cdot 2^l\} \right) $$
The support of the interpolet on level $0$ at $y$ is $[-3+y,3+y]$, the
union of all the supports of the interpolets in $S$ on level $0$ is
$[-10,10]$.  The support of an interpolet on level $1$ at $y$ is
$[-6+y,6+y]$, thus the only interpolets on level $1$ which touch the
interpolets on level $0$ are precisely those ones at points
$-14,\dots, 14$, which are precisely the ones included in $S$.  No
interpolet not included on level 1 touches an interpolet included on
level 0 so the definition is satisfied.  The argument proceeds
similarly on higher levels.  In three dimensions, this example
corresponds to nested concentric cubes of size $15^3 \cdots 2^l$ at each
level $l<n$.

An example of a 2-level touching good basis for 3rd order is the set
of interpolets centered at the points 
$$ S = C_n \cup \left( \bigcup^{n-1}_{l=0} \{-5\cdot 2^l,-3\cdot 2^l,\dots,3\cdot 2^l,5\cdot 2^l\} \right).$$
Note (for $n\ge2$) that this set is not 1-level touching since the
level 1 interpolet centered at $y=12$ is not included, while an interpolet
it touches, namely the level 0 interpolet centered at $y=5$, is included.

Also, the set of points 
$$ S = C_n \cup \left( \bigcup^{n-1}_{l=0} \{-3\cdot 2^l,-1\cdot 2^l,1\cdot 2^l,3\cdot 2^l\} \right).$$
forms a good basis, but is not 1-level touching or 2-level
touching (it is 3-level touching, though).

The above examples are meant to suggest that the {\em good basis} and
the {\em p-level touching} definitions can be thought of, informally,
as conditions telling one how quickly one can change from one
resolution to another.  Essentially, any nested set of refined regions
can satisfy these conditions so long as the margins around the set of
points at a given resolution are wide enough.

>From a computational point of view, what these conditions do is ensure
that data-paths which carry coefficient information between different
resolutions are not broken by zeroed coefficients at intermediate
levels.

It is clear from the preceding discussion, in a good basis,
one has $J_{k_1, k_2} = J_{k_2-1, k_2} \circ J_{k_2-2, k_2-1} \circ
\cdots \circ J_{k_1, k_1+1}.$ We may now generalize the computation
theorem to a truncated basis.

\begin{theorem}{Good Basis Computation Theorem\\}
Let $S$ be a good basis.
$v \in \curlyF_k(S)$, $y \in S$\\
and $\tilde{v}$ is any member of the equivalence class of $v$\\
$$(J_{k, k+1} \tilde{v})(y) = \left\{ \begin{array}{lc}
v(y), & y \in S - D_{k+1} \\  v'(y), & y \in D_{k+1}
\end{array}\right.$$
where $v'(y) = v(y) - \sum_{z \in S\cap C_{k+1}} c_{(y-z)/2^k} v(z).$

$$(J_{k+1, k} \tilde{v})(y) = \left\{ \begin{array}{lc}
v(y), & y \in S - D_{k+1} \\  v'(y), & y \in D_{k+1}
\end{array}\right.$$

where $v'(y) = v(y) + \sum_{z \in S\cap C_{k+1}} c_{(y+z)/2^k} v(z).$

\end{theorem} 

\proof

In a good basis, the computations of all $J_{k_1, k_2}$ are
independent of the representative.  Thus, this algorithm computed on
$\tilde{v}$ gives a member of the same class as would be computed
on $v$.

$\Box$

Thus, the pyramid algorithm of the uniform basis, $Z^n$, has
a counterpart in a good basis $S$, allowing the computation of
the expansion coefficients in $\curlyF_k(S)$ from the values of
the function in $\curlyF_0(S)$ (and also has the lower triangular
structure).

In a good basis, one thus has the ability to perfectly reconstruct the
multiscale coefficients of a function for the basis functions
associated with the points of the refined grid $S$ by simply applying
the pyramid algorithm on zero-lifts at each stage of the algorithm.
The above theorem establishes this as true, even though we do not
necessarily expect the data zeroed during the lift to be small.  (The
function may have significant sample values throughout the domain of
the representation).  Also, with exact recovery of sample values, it
is easy to perform local nonlinear point-wise operations of the form
$f(x) \rightarrow G(f(x))$ (e.g. $e^{f(x)}$), or point-wise
multiplication (i.e. $f(x), g(x) \rightarrow f(x)g(x)$), despite the
truncation.

The reader may note that this result is "analysis-free" in that we
have sparsified the computation, not by proving that the coefficients
vanish outside of the truncation for some class of functions, but by
showing the the coefficients we wish to compute have no dependency on
the coefficients we omitted.  Computationally, this means the
data-structure in the computer requires no intermediate augmentations
(contrast with \cite{GoedeckerIvanov:97}).

The advantages conferred by the additional the $p$-level properties are
seen in the context of operator evaluation, and will be the subject of
the next two sections.

\section{Multilevel Algorithms for $\nabla^2$ and Other Operators}

Given $v,w \in \curlyF_k(Z^n)$, we may compute the stiffness
matrix of a model operator, $\nabla^2$,
$$\left<\iota_k v |\nabla^2 | \iota_k w\right> = \int (\iota_k v)(x)\nabla^2 (\iota_k w)(x) d^nx$$
by changing the expansions,
$$\left<\iota_0 J_{-k} v |\nabla^2 | \iota_0 J_{-k} w\right> = 
\sum_{y,z \in Z^n} (J_{-k} v)(y) (J_{-k} w)(z) \int \phi(x-y) \nabla^2 \phi(x-z) d^nx.$$
This reduces the computation to the computation of the matrix elements
\hbox{$\left<\phi(x-(y-z))|\nabla^2|\phi(x)\right>$} which can be done by solving the 
associated eigen-problem obtained by applying (INT2).

In particular, let $L^0_y = \left<\phi(x-y)|\nabla^2|\phi(x)\right>$, then
$$L^0_y = \sum_{z_1,z_2\in Z^n} c_{z_1} c_{z_2} \left<\phi(2x-2y-z_1+z_2)|\nabla^2|\phi(2x)\right>$$
$$L^0_y = \sum_{z_1,z_2\in Z^n} 2^{2-n} c_{z_1} c_{z_2} L^0_{2y+z_1-z_2}$$
which we solve by standard methods (found in \cite{Strang.book:96},
for example).  In subsequent sections of this article, 
we will also
define 
$L^{+}_y = \left<\phi(x-y)|\nabla^2|\phi(x/2)\right>$,
$L^{++}_y = \left<\phi(x-y)|\nabla^2|\phi(x/4)\right>$,
$L^{-}_y = \left<\phi((x-y)/2)|\nabla^2|\phi(x)\right>$,
 and 
$L^{--}_y = \left<\phi((x-y)/4)|\nabla^2|\phi(x)\right>$, which can be 
computed from $L^0$ by employing (INT2).
Although it is true by Hermiticity that 
$L^{-}_y = L^{+}_y$ and $L^{--}_y = L^{++}_y$, we will make no use
of this fact.

One can write a matrix expression $\left<\iota_k v|\nabla^2 | \iota_k w\right> = 
v^t J_{-k}^t L J_{-k} w,$ where the $L$ is
the Toeplitz matrix with coefficients $L^0_y$.
In practice, one typically formulates the preceding
as a computation of $u = J_{-k}^t D J_{-k} w$ for some $w \in
\curlyF_k(Z^n)$.  Then \hbox{$\left<\iota_k v|\nabla^2 | \iota_k w\right>$}$ = v^t u$.
$u$ may be thought of as an element of $\curlyF_k(Z^n)^*$, the dual
space of $\curlyF_k(Z^n)$.  The task is then to compute the
coefficients $u(y), y\in Z^n = C_k \oplus D_k \oplus \cdots \oplus
D_1$.  Any algorithm for computing $x^tAy$ can be adapted to an
algorithm for $Ay$, and for purposes of making proofs, it is somewhat
easier to keep thinking of the computation as \hbox{$\left<\iota_k v | \nabla^2 |
\iota_k w\right>$}, which is the point of view we shall take.



However, computing \hbox{$\left<\iota_k \tilde{v} | \nabla^2 | \iota_k \tilde{w}\right>$}$ =
\tilde{v}^t J_{-k}^t L J_{-k} \tilde{w},$ by just applying the
transforms, and the Toeplitz matrix is problematic, since
this process makes it necessary to either
represent $v$ and $w$ on a uniform grid, or to compute
a matrix element between each pair of functions in the truncated
expansion which touch.  In the first case, one ends up with an $O(N)$
computation for (typically) a very large $N$.  In the second case, one
chooses between extremes which are $O(N)$ and quite complicated or
simple and $O(N^2)$.

The following sections outline the design of multilevel algorithms
for $\nabla^2$ for both 1-level and 2-level touching bases.  Both
algorithms are derived according to the following format:

\begin{list}{}
\item break up the expansion of \hbox{$\left<\iota_k v| \nabla^2 |\iota_k w\right>$}
into a decomposition over elements at the same level and adjacent levels.
\item rewrite the expansions in terms of the matrix elements between elements 
of those levels and the transforms of higher/lower level elements.
\item implement the algorithm by computing those terms separately.
\item establish correctness in a $p$-level truncated basis.
\end{list}

Although only the 1-level and 2-level algorithms have been explored
in any detail, this same process will generally work to produce
$O(N)$ $p$-level algorithms for any $p$.

\subsection{$\nabla^2$ in 1-level decomposition}

The 1-level decomposition of \hbox{$\left<\iota_k v | \nabla^2 | \iota_k w\right>$} is
$$\left<\iota_k v | \nabla^2 | \iota_k w\right> = \left<\iota_k v | \nabla^2 | \iota_k w\right>^+ + \left<\iota_k v | \nabla^2 | \iota_k w\right>^0 + \left<\iota_k v | \nabla^2 | \iota_k w\right>^-,$$
where
\begin{eqnarray*}
\left<\iota_k v | \nabla^2 | \iota_k w\right>^0 = \sum_{\level_k(y) = \level_k(z)} \left<\iota_k y |\nabla^2 | \iota_k z \right> v(y)w(z)\\
\left<\iota_k v | \nabla^2 | \iota_k w\right>^+ = \sum_{\level_k(y) < \level_k(z)} \left<\iota_k y |\nabla^2 | \iota_k z \right> v(y)w(z)\\
\left<\iota_k v | \nabla^2 | \iota_k w\right>^- = \sum_{\level_k(y) > \level_k(z)} \left<\iota_k y |\nabla^2 | \iota_k z \right> v(y)w(z).
\end{eqnarray*}
That is, we express the product as contributions from levels to the
same level, higher levels, and lower levels.  We will now investigate
each of these terms individually.  In the language of matrices, these
respectively
correspond to diagonal, above diagonal, and below diagonal blocks
of the $\nabla^2$ matrix.

$$ \left<\iota_k v |\nabla^2| \iota_k w\right> = 
v^t \pmatrix{\left<|\right>^0 & \left<|\right>^+ & \left<|\right>^+ & \left<|\right>^+ \cr 
	 \left<|\right>^- & \left<|\right>^0 & \left<|\right>^+ & \left<|\right>^+ \cr 
	 \left<|\right>^- & \left<|\right>^- & \left<|\right>^0 & \left<|\right>^+ \cr 
	 \left<|\right>^- & \left<|\right>^- & \left<|\right>^- & \left<|\right>^0 \cr} w $$

\subsubsection{Diagonal blocks: \hbox{$\left<|\right>^0$}}
For some fixed $l, 0\le l\le k$ a term of the form
$$\sum_{\level_k(z) = \level_k(y) = l} \left<\phi((x-y)/2^l) | \nabla^2 | \phi((x-z)/2^l)\right> v(y)w(z)$$
contributes to \hbox{$\left<|\right>^0$}.

However, \hbox{$\left<\phi((x-y)/2^l) | \nabla^2 | \phi((x-z)/2^l)\right>$} $= 2^{l (n-2)}
\left<\phi(x-y)| \nabla^2 | \phi(x-z)\right> = 2^{l (n-2)} L^0_{z-y}$.  Thus we have
$$\left<\iota_k v| \nabla^2 | \iota_k w\right>^0 = \sum_{l} 2^{l (n-2)} \sum L^0_{z-y} v(y)w(z).$$

\subsubsection{Super-diagonal blocks: \hbox{$\left<|\right>^+$}}
For some fixed $l, 0\le l\le k$ a term of the form
$$\sum_{\level_k(z)=l' >l=\level_k(y)} \left<\phi((x-y)/2^l) | \nabla^2 | \phi((x-z)/2^{l'})\right> v(y)w(z)$$
contributes to \hbox{$\left<|\right>^+$}.

Applying the inverse transform ($J_{k, l+1}$) to the $w$ coefficients
in the sum allows us to write this term as
$$\sum_{\level_{l+1} (z)=l+1 >l = \level_k (y)} \left<\phi((x-y)/2^l) | \nabla^2 | \phi((x-z)/2^{l+1})\right> v(y) (J_{k, l+1} w)(z).$$

Thus we have
$$\left<\iota_k v| \nabla^2| \iota_k w\right>^+ = 
\sum_{l} 2^{l (n-2)} \sum L^{+}_{z-y} v(y)(J_{k, l+1} w)(z).$$

\subsubsection{Sub-diagonal blocks: \hbox{$\left<|\right>^-$}}
For some fixed $l, 0\le l\le k$ a term of the form
$$\sum_{\level_k(z)=l <l'=\level_k(y)} \left<\phi((x-y)/2^{l'}) | \nabla^2 | \phi((x-z)/2^{l})\right> v(y)w(z)$$
contributes to \hbox{$\left<|\right>^-$}.

This time applying the inverse transform ($J_{k, l+1}$) to the $v$ coefficients
in the sum allows us to write this term as
$$\sum_{\level_k (z)=l <l+1 = \level_{l+1} (y)} \left<\phi((x-y)/2^{l+1}) | \nabla^2 | \phi((x-z)/2^l)\right> (J_{k, l+1} v)(y) w(z).$$

Thus we have
$$\left<\iota_k v| \nabla^2| \iota_k w\right>^- = 
\sum_{l} 2^{l (n-2)} \sum L^{-}_{z-y} (J_{k, l+1} v)(y)w(z).$$

\subsubsection{Implementations}

The above observations demonstrate the correctness of the following algorithm:\\
input: $v,w \in \curlyF_k(Z^n)$\\
output: $ans = \left<\iota_k v | \nabla^2 | \iota_k w\right> \in R$\\
\indent Let $wtmp = w$, let $vtmp = v$, let $ans = 0$\\ 
\indent for $l=0$ to $k$\\
\indent \hspace{0.5 in} $ans \leftarrow ans + 2^{l(n-2)}\sum_{\level_k(y)=\level_k(z)=l} L^0_{z-y} v(y)w(z)$\\
\indent end for\\
\indent for $l=k-1$ down-to $0$\\
\indent \hspace{0.5 in} $ans \leftarrow ans + 2^{l(n-2)}\sum_{\level_k(y)=l,\level_{l+1}(z)=l+1} L^{+}_{z-y} v(y)wtmp(z)$\\
\indent \hspace{0.5 in} $wtmp \leftarrow J_{l+1, l} (wtmp)$\\
\indent end for\\
\indent for $l=k-1$ down-to $0$\\
\indent \hspace{0.5 in} $ans \leftarrow ans + 2^{l(n-2)}\sum_{\level_{l+1}(y)=l+1,\level_k(z)=l} L^{-}_{z-y} vtmp(y)w(z)$\\
\indent \hspace{0.5 in} $vtmp \leftarrow J_{l+1, l} (vtmp)$\\
\indent end for\\
Note, we have made use of the fact that $J_{k, l} = J_{l+1,
 l} \cdots J_{k, k-1}$
so that at the beginning of each iteration in the second (last) loop $wtmp = J_{k, l+1} w$ ($vtmp = J_{k, l+1} v$).

We observe that this algorithm is $O(N)$ in time and space. 

We may adapt this to an
algorithm to compute $u \in \curlyF_k(Z^n)^*$ such that $u(v) =$
\hbox{$\left<\iota_k v | \nabla^2 | \iota_k w\right>.$}\\
input: $w \in \curlyF_k(Z^n)$\\
output: $u$ such that $u(v) =$ \hbox{$\left<\iota_k v | \nabla^2 | \iota_k w\right>$}\\
\indent Let $wtmp = w$, let $u = 0 \in \curlyF_k(Z^n)^*$, let $utmp = 0 \in \curlyF_0(Z^n)^*$\\
\indent for $l=0$ to $k$\\
\indent \hspace{0.5 in} $u(y) \leftarrow u(y) + 2^{l(n-2)}\sum_{\level_k(y)=level_k(z)=l} L^0_{z-y} w(z)$\\
\indent end for\\
\indent for $l=k-1$ down-to $0$\\
\indent \hspace{0.5 in} $u(y) \leftarrow u(y) + 2^{l(n-2)}\sum_{\level_k(y)=l,level_{l+1}(z)=l+1} L^{+}_{z-y} wtmp(z)$\\
\indent \hspace{0.5 in} $wtmp \leftarrow J_{l+1, l} wtmp$\\
\indent end for\\
\indent for $l=0$ to $k-1$\\
\indent \hspace{0.5 in} $utmp \leftarrow J_{l+1, l}^t utmp$\\
\indent \hspace{0.5 in} $utmp(y) \leftarrow utmp(y) + 2^{l(n-2)}\sum_{\level_{l+1}(y)=l+1,\level_k(z)=l} L^{-}_{z-y} w(z)$\\
\indent end for\\
\indent $u \leftarrow u + utmp$

The third loop is the result of transposing the linear operator $J_{l+1,l}$
in the last loop in the previous algorithm.  We have also used the fact that
$J_{k, l}^t = J_{k, k-1}^t \cdots J_{l+1, l}^t$ to ensure that at the
beginning of each iteration in the third loop, $utmp \in
\curlyF_{l+1}(Z^n)^*$.


It is easy to check that this is an $O(N)$ algorithm in both time and
space.  It is very important for atomic structures computation that
this algorithm scales linearly with the number of atoms.  Without such
a scaling, one can only compute electronic configurations for small
molecules.

The reader may note a similarity between this algorithm and other
matrix-vector multiplies used to apply operators in a uniform wavelet
basis.  In fact, the 1-level algorithm presented above is identical to
the nonstandard multiply found in \cite{BeylkinCoifmanRokhlin:91} and
developed for orthonormal wavelet bases.  The nonstandard multiply was
introduced by Beylkin, Coifman, and Rokhlin to sparsify integral
operators whose kernels were smooth or vanishing off the diagonal,
while keeping a uniform basis.

However, in contrast to that program of sparsification, interpolets
allow one to sparsify the basis, and, with out introducing additional
grid points, still be able to apply the nonstandard multiply routines,
with any local operator.  With interpolets, we remove any elements
from the expansion that we believe will be insignificant, still having
a good approximation to our function at the points we retain.  Beylkin
{\em et al.} \cite{BeylkinCoifmanRokhlin:91} express the matrix
elements of the operator itself in a nonstandard orthonormal basis and
then remove those matrix elements which are determined to be very
small to produce a sparse matrix.

The use of interpolating scaling functions has achieved some degree of
simplicity and convenience in carrying out fast point-wise operations.
Although there is no associated difficulty in electronic structure
calculations\cite{Arias.unpublished:97}, for other applications, the
loss of orthogonality might be too great an expense.  In those cases,
one might consider employing compactly supported approximations to
orthogonal interpolating functions found in \cite{BeylkinKeiser:95}.
It appears that with some additional complexity one might be able to
extend the present algorithms to other wavelet bases.
The additional complexity of other schemes and the need for fast
point-wise operations in our applications are the chief reasons we do
not consider doing this in the present work.  Finally, there is a
large body of work the reader may wish to consult
(\cite{BeylkinKeiser:97}, \cite{FrohlichSchneider:94},
\cite{DahlkeDahmenHochmuthSchneider:97}, \cite{liandrat}, and
\cite{LiandratTchamitchian:90}) for adaptive refinement techniques
when, in contrast to the case of electronic structure calculations,
the behavior of the needed refinement is not known {\em a priori}.

\subsubsection{Correctness in a 1-level touching good basis}

The above decomposition of the product and the associated algorithm
is what we seek to extend to a good truncated
basis.  In practice, one takes the zero-lift representatives of $v$
and $w$ $\in
\curlyF^S_k$ and computes \hbox{$\left<\iota_k \tilde{v}| \nabla^2 |\iota_k
\tilde{w}\right>$}.  By the computation theorem of good truncated bases, the
value of $(J_{k_1, k_2}\tilde{v})(y), y \in
\curlyF^S_k$ is independent of the representative (likewise for $w$), 
however, we must also address the issue that $(J_{k_1,
k_2}\tilde{v})(y) \ne 0, y \notin S$ 
(i.e. $J_{k_1, k_2} \tilde{v} \ne \widetilde{J_{k_1, k_2} v}$),
and thus $y \notin S$ may have a contribution to the
decomposition above, requiring us to augment $S$ in order to
get the right answer.

\begin{theorem}{}
If one replaces $Z^n$ with $S$ everywhere in the Multilevel algorithm for
\hbox{$\left<\iota_k v| \nabla^2 |\iota_k w\right>$}, then the algorithm computes 
$$\left<\iota_k \tilde{v}| \nabla^2 |\iota_k \tilde{w}\right>.$$
\end{theorem}

\proof
As mentioned in the remarks, the multilevel algorithm requires
$J_{k, l} = J_{l+1, l} \cdots J_{k, k-1}$, which is true in a good
bases.

The \hbox{$\left<|\right>^0$} computation proceeds identically for either $Z^n$ or $S$,
so the first loop will contribute correctly the term \hbox{$\left<\iota_k
\tilde{v} | \nabla^2 | \iota_k \tilde{w}\right>^0$}.

To check the \hbox{$\left<|\right>^+$} contribution, one observes that the necessary term is
$$\left<\iota_k \tilde{v}| \nabla^2| \iota_k \tilde{w}\right>^- = 
\sum_{l} 2^{l (n-2)} \sum_{y\in Z^n, z\in Z^n:\level_{l+1}(y)=l+1,\level_k(z)=l} L^{-}_{z-y} (J_{k, l+1} \tilde{v})(y)\tilde{w}(z).$$

Suppose that $\exists y \notin S$ and $z \in S$ such that $L^{+}_{z-y}
\ne 0$.  This implies that $\supp\{\iota_{l+1} y\} \cap
\supp\{\iota_k z\}$ has nonzero measure.  Since $\supp\{\iota_{l+1} y\}
\subset \supp\{\iota_k y\}$ we conclude that $\supp\{\iota_k y\} 
\cap \supp\{\iota_k z\}$ has nonzero measure.  This cannot be so if $S$ is 1-level touching, and since $\tilde{w}(z) = 0, z\notin S$ we may
restrict the sums over $y$ and $z$ in the contribution,
$$2^{l (n-2)} \sum_{y\in S, z\in S:\level_{l+1}(y)=l+1,\level_k(z)=l} L^{-}_{z-y} (J_{k, l+1} \tilde{v})(y)\tilde{w}(z).$$

The proof for \hbox{$\left<|\right>^-$} is identical with $v$'s and $w$'s reversed.

$\Box$


Immediately we have the following:
\begin{corollary}{}
If one replaces $Z^n$ with $S$ everywhere in the Multilevel algorithm
for $u$ such that $u(v) =$ \hbox{$\left<\iota_k v| \nabla^2 |\iota_k w\right>$},
then the algorithm computes 
$$u \in (\curlyF^S_k)^*, u(v) = \left<\iota_k \tilde{v}| \nabla^2|\iota_k \tilde{w}\right>.$$
\end{corollary}

The computation of \hbox{$\left<\iota_k v| \nabla^2 | \iota_k w\right>$} serves as
a template for another common computation one may wish to perform,
namely \hbox{$\left<\iota_k v| \iota_k w\right>$} $ = \int (\iota_k v)(x)(\iota_k
w)(x) d^n x$, i.e. the $L^2(R^n)$ inner product of $\iota_k v$ and
$\iota_k w$.  

\subsection{Computing Other Operators}

To compute \hbox{$\left<\iota_k \tilde{v}| \iota_k \tilde{w}\right>$}, one simply
replaces $L^0$, $L^{+}$, and $L^{-}$ with $G^0_y =$
\hbox{$\left<\phi(x-y)|\phi(x)\right>$}, $G^+_y =$
\hbox{$\left<\phi(x-y)|\phi(x/2)\right>$}, and $G^-_y =$
\hbox{$\left<\phi((x-y)/2)|\phi(x)\right>$}, and then replaces the factors of
$2^{l(n-2)}$ with $2^{ln}$.  After that, the algorithms and theorems for
$\nabla^2$ carry over directly.

The above procedure can be used for creating a multilevel algorithm
for any operator, ${\cal O}$, which is 
\begin{list}{}
\item local :$\supp\{ {\cal O} f \} \subset \supp \{ f\} $
\item translation invariant :${\cal O}f(x+a) = ({\cal O} f)(x+a)$
\item homogeneous :${\cal O} f(sx) = s^\alpha ({\cal O} f)(sx)$
\end{list}
\noindent
by forming the appropriate coefficients, ${\cal O}^{ \{0,+,-\} }_y$, and
inserting appropriate factors of $2^{l(\alpha+d)}$.

However, locality is the only property which is really required for
$O(N)$ multilevel algorithms so long as one can compute ${\cal
O}^{\{0,+,-\}}_{l,m,m'}$ efficiently.

As an example of a local, homogeneous, but not translationally
invariant operator, we shall discuss the coefficients for the
multilevel algorithm for the $\hat{x_1}$ operator in two dimensions.

We first consider the coefficients $\hat{x_1}^{\{0,+,-\} }_{l,m,m'}$ given
by 
$$\hat{x_1}^{\{0,+,- \} }_{l,m,m'} = \int \phi((x_1-m_1)/2^{l(+1)})\phi((x_2-m_2)/2^{l(+1)}) x_1 \phi((x_1-m'_1)/2^{l(+1)})\phi((x_2-m'_2)/2^{l(+1)}) dx_1dx_2,$$
i.e.
\begin{eqnarray*}
\hat{x_1}^0_{l,m,m'} &=& \int \phi((x_1-m_1)/2^l)\phi((x_2-m_2)/2^l) x_1 \phi((x_1-m'_1)/2^l)\phi((x_2-m'_2)/2^l) dx_1dx_2 \cr
\hat{x_1}^+_{l,m,m'} &=& \int \phi((x_1-m_1)/2^l)\phi((x_2-m_2)/2^l) x_1 \phi((x_1-m'_1)/2^{l+1})\phi((x_2-m'_2)/2^{l+1}) dx_1dx_2 \cr
\hat{x_1}^-_{l,m,m'} &=& \int \phi((x_1-m_1)/2^{l+1})\phi((x_2-m_2)/2^{l+1}) x_1 \phi((x_1-m'_1)/2^l)\phi((x_2-m'_2)/2^l) dx_1dx_2.  \cr
\end{eqnarray*}

Separating the $x_1$ and $x_2$ integrations, we see from this that we
may write $\hat{x_1}^{\{0,+,-\} }_{l,m,n} = 2^{l(d-1)} G^{\{0,+,-\}
}_{(m_1-n_1)/2^l} X^{\{0,+,-\} }_{l,m_2,n_2}$ 
where $$X^{\{0,+,-\} }_{l,m,n} = \int \phi((x-m)/2^{l(+1)}) x
\phi((x-n)/2^{l(+1)}) dx$$ and $G$ is defined above.  The problem of
computing the $\hat{x_1}$ coefficients has been reduced to computing
the $X$ coefficients and then doing a multiply with the already known
$G$ coefficients.

In addition, one has 
$$\int \phi((x-m)/2^{l(+1)}) x \phi((x-n)/2^{l(+1)}) dx = 
n \int \phi((x+n-m)/2^{l(+1)}) \phi(x/2^{l(+1)}) dx + $$ $$\int \phi((x+n-m)/2^{l(+1)})x \phi(x/2^{l(+1)}) dx,$$
and thus
$$\int \phi((x-m)/2^{l(+1)}) x \phi((x-n)/2^{l(+1)}) dx = 
n 2^{l} G^{\{0,+,-\} }_{(m-n)/2^l} + 2^{2l} S^{\{0,+,-\} }_{(m-n)/2^l}$$
where
\begin{eqnarray*}
S^0_y &=& \int \phi(x-y) x \phi(x) dx \cr
S^+_y &=& \int \phi(x-y) x \phi(x/2) dx \cr
S^-_y &=& \int \phi((x-y)/2) x \phi(x) dx. \cr
\end{eqnarray*}

Thus we see that for the $\hat{x_1}$ operator,
$$\hat{x_1}^{\{0,+,-\} }_{l,m,n} = 2^l G^{\{0,+,-\} }_{(m_2-n_2)/2^l} (n 2^l
G^{\{0,+,-\} }_{(m-n)/2^l} + 2^{2l} S^{\{0,+,-\} }_{m-n}),$$
giving an efficient means to compute the multilevel coefficient
for this operator.

\subsection{$\nabla^2$ in 2-level decomposition}

The previous algorithm for 1-level touching good bases can be expanded
to 2-level touching good bases.  One may wish to do this because one
finds that the 1-level touching property is too stringent and requires
one to augment one's basis set far too much to be practical
computationally.

Much of the reasoning for the 2-level case can be found in the details
of the 1-level case, so the exposition here will be more compact.  The
resulting algorithm will be correct for 2-level touching good bases.

The 2-level decomposition of \hbox{$\left<\iota_k v | \nabla^2 | \iota_k w\right>$} is
$$\hbox{$\left<\iota_k v | \nabla^2 | \iota_k w\right>$} = \hbox{$\left<\iota_k v |
\nabla^2 | \iota_k w\right>^0$} + \hbox{$\left<\iota_k v | \nabla^2 | \iota_k
w\right>^+$} + \hbox{$\left<\iota_k v | \nabla^2 | \iota_k w\right>^-$}$$ $$ +
\hbox{$\left<\iota_k v | \nabla^2 | \iota_k w\right>^{++}$} +\hbox{$\left<\iota_k v |
\nabla^2 | \iota_k w\right>^{--}$}$$

where
\begin{eqnarray*}
\left<\iota_k v | \nabla^2 | \iota_k w\right>^0 = \sum_{\level_k(y) = \level_k(z)} \left<\iota_k y |\nabla^2 | \iota_k z\right> v(y)w(z)\\
\left<\iota_k v | \nabla^2 | \iota_k w\right>^+ = \sum_{\level_k(y)+1 = \level_k(z)} \left<\iota_k y |\nabla^2 | \iota_k z \right> v(y)w(z)\\
\left<\iota_k v | \nabla^2 | \iota_k w\right>^- = \sum_{\level_k(y)-1 = \level_k(z)} \left<\iota_k y |\nabla^2 | \iota_k z \right> v(y)w(z)\\
\left<\iota_k v | \nabla^2 | \iota_k w\right>^{++} = \sum_{\level_k(y)+1 < \level_k(z)} \left<\iota_k y |\nabla^2 | \iota_k z \right> v(y)w(z)\\
\left<\iota_k v | \nabla^2 | \iota_k w\right>^{--} = \sum_{\level_k(y)-1 > \level_k(z)} \left<\iota_k y |\nabla^2 | \iota_k z \right> v(y)w(z)
\end{eqnarray*}

Which corresponds to the matrix decomposition:
$$ \left<\iota_k v |\nabla^2| \iota_k w\right> = 
v^t \pmatrix{\left<|\right>^0 & \left<|\right>^+ & \left<|\right>^{++} & \left<|\right>^{++} \cr 
	 \left<|\right>^- & \left<|\right>^0 & \left<|\right>^+ & \left<|\right>^{++} \cr 
	 \left<|\right>^{--} & \left<|\right>^- & \left<|\right>^0 & \left<|\right>^+ \cr 
	 \left<|\right>^{--} & \left<|\right>^{--} & \left<|\right>^- & \left<|\right>^0 \cr} w. $$

The key idea is to evaluate the diagonal and first off diagonal
blocks of the $\nabla^2$ matrix and then to compute the other blocks
above and below the tridiagonal through the transforms.

\subsubsection{\hbox{$\left<|\right>^0$}, \hbox{$\left<|\right>^+$}, \hbox{$\left<|\right>^-$}, \hbox{$\left<|\right>^{++}$}, and \hbox{$\left<|\right>^{--}$}} 

The definitions for the contributions in the decomposition proceed just as
they did for the 1-level case.

$$\left<\iota_k v| \nabla^2 | \iota_k w\right>^0 = 
\sum_{l} 2^{l (n-2)} \sum_{\level_k(z) = \level_k(y) = l} L^0_{z-y} v(y)w(z).$$

$$\left<\iota_k v| \nabla^2| \iota_k w\right>^+ = 
\sum_{l} 2^{l (n-2)} \sum_{\level_k(z)-1 = \level_k(y) = l} L^{+}_{z-y} v(y)w(z).$$

$$\left<\iota_k v| \nabla^2| \iota_k w\right>^- = 
\sum_{l} 2^{l (n-2)} \sum_{\level_k(z) = \level_k(y)-1 = l} L^{-}_{z-y} v(y)w(z).$$

$$\left<\iota_k v| \nabla^2| \iota_k w\right>^{++} = 
\sum_{l} 2^{l (n-2)} \sum_{\level_k(z)-1 >l = \level_k(y)} L^{++}_{z-y} v(y)(J_{k, l+2} w)(z).$$

$$\left<\iota_k v| \nabla^2| \iota_k w\right>^{--} = 
\sum_{l} 2^{l (n-2)} \sum_{\level_k(z) = l < \level_k(y)-1} L^{--}_{z-y} (J_{k, l+2} v)(y)w(z).$$

\subsubsection{Implementations}

input: $v,w \in \curlyF_k(Z^n)$\\
output: $ans = \left<\iota_k v | \nabla^2 | \iota_k w\right> \in R$\\
\indent Let $wtmp = w$, let $vtmp = v$, let $ans = 0$\\ 
\indent for $l=0$ to $k$\\
\indent \hspace{0.5 in} $ans \leftarrow ans + 2^{l(n-2)}\sum_{\level_k(y)=\level_k(z)=l} L^0_{z-y} v(y)w(z)$\\
\indent end for\\
\indent for $l=0$ to $k-1$\\
\indent \hspace{0.5 in} $ans \leftarrow ans + 2^{l(n-2)}\sum_{\level_k(y)=l,\level_k(z)=l+1} L^{+}_{z-y} v(y)w(z)$\\
\indent end for\\
\indent for $l=0$ to $k-1$\\
\indent \hspace{0.5 in} $ans \leftarrow ans + 2^{l(n-2)}\sum_{\level_k(y)=l+1,\level_k(z)=l} L^{-}_{z-y} v(y)w(z)$\\
\indent end for\\
\indent for $l=k-2$ down-to $0$\\
\indent \hspace{0.5 in} $ans \leftarrow ans + 2^{l(n-2)}\sum_{\level_k(y)=l,\level_{l+2}(z)=l+2} L^{++}_{z-y} v(y)wtmp(z)$\\
\indent \hspace{0.5 in} $wtmp \leftarrow J_{l+1, l} wtmp$\\
\indent end for\\
\indent for $l=k-2$ down-to $0$\\
\indent \hspace{0.5 in} $ans \leftarrow ans + 2^{l(n-2)}\sum_{\level_{l+2}(y)=l+2,\level_k(z)=l} L^{--}_{z-y} vtmp(y)w(z)$\\
\indent \hspace{0.5 in} $vtmp \leftarrow J_{l+1, l} vtmp$\\
\indent end for\\

We adapt this
algorithm to compute $u \in \curlyF_k(Z^n)^*$ such that $u(v) =$
\hbox{$\left<\iota_k v | \nabla^2 | \iota_k w\right>.$}\\
input: $w \in \curlyF_k(Z^n)$\\
output: $u$ such that $u(v) =$ \hbox{$\left<\iota_k v | \nabla^2 | \iota_k w\right>$}\\
\indent Let $wtmp = w$, let $u = 0 \in \curlyF_k(Z^n)^*$, let $utmp = 0 \in \curlyF_0(Z^n)^*$\\
\indent for $l=0$ to $k$\\
\indent \hspace{0.5 in} $u(y) \leftarrow u(y) + 2^{l(n-2)}\sum_{\level_k(y)=\level_k(z)=l} L^0_{z-y} w(z)$\\
\indent end for\\
\indent for $l=0$ to $k-1$\\
\indent \hspace{0.5 in} $u(y) \leftarrow u(y) + 2^{l(n-2)}\sum_{\level_k(y)=l,\level_k(z)=l+1} L^{+}_{z-y} w(z)$\\
\indent end for\\
\indent for $l=0$ to $k-1$\\
\indent \hspace{0.5 in} $u(y) \leftarrow u(y) + 2^{l(n-2)}\sum_{\level_k(y)=l+1,\level_k(z)=l} L^{-}_{z-y} w(z)$\\
\indent end for\\
\indent for $l=k-2$ down-to $0$\\
\indent \hspace{0.5 in} $u(y) \leftarrow u(y) + 2^{l(n-2)}\sum_{\level_k(y)=l,\level_{l+2}(z)=l+2} L^{++}_{z-y} wtmp(z)$\\
\indent \hspace{0.5 in} $wtmp \leftarrow J_{l+1, l} wtmp$\\
\indent end for\\
\indent for $l=0$ to $k-2$\\
\indent \hspace{0.5 in} $utmp \leftarrow J_{l+1, l}^t utmp$\\
\indent \hspace{0.5 in} $utmp(y) \leftarrow utmp(y) + 2^{l(n-2)}\sum_{\level_{l+2}(y)=l+2,\level_k(z)=l} L^{--}_{z-y} w(z)$\\
\indent end for\\
\indent $u \leftarrow u + utmp$

\section{Efficient Implementation}

We have produced a very successful 3D implementation of all of the
above algorithms for the interpolet used as this paper's example
($m=3$).  Implementation details are given in this
section.  The ideas used to make this implementation efficient
for all of the above algorithms.

The purpose of this section is to give additional information to 
readers who wish to implement these algorithms themselves.

\subsection{Data Structures}

The interpolet data and function samples are kept in a sequence of
blocks at various levels.  Each block at level $k$ contains the points
of a rectangular subset of $C_{k-1}$.  Since $D_k = C_{k-1} - C_{k}$,
we use the collection of blocks at level $k < p$ ($p$ being the top
level) to represent a rectangular subset $O_k$, ignoring the $C_{k}$
points of each of these blocks.  In our implementation, these extra
$C_k$ points hold the value $0$ in between operations and take on
useful intermediate values during operations.  Since we are working in
3 dimensions, this multiplies the storage required by a factor of
about $\frac{8}{7}$, which we found an acceptable cost for its
advantages.

The coefficients for the transforms and operators are kept in various
3D arrays.  Although it is possible to build the coefficients upon
demand from a set of 1D arrays of coefficients, we have found that the
arithmetic cost of doing this is much greater than the cost of storing them
(about 10 flops are required for each $\nabla^2$ coefficient, while the
3D arrays are still small enough to be stored in cache).  We have (in
Fortran notation) the filters cs(0:3,0:3,0:3), SAMELEVEL(0:5,0:5,0:5),
ONELEVEL(0:8,0:8,0:8), and (for 2-level algorithms)
TWOLEVEL(0:14,0:14,0:14) (note: we have made use of the fact that
our operators are symmetric to cut the size of these arrays by a factor of
$\frac{1}{8}$ and use ONELEVEL and TWOLEVEL for both upward and
downward inter-level communication).

\subsection{Implementation Overview}

The blocks described in the previous section are used as the fundamental
objects for manipulation.  The computation proceeds by employing
block-to-block subroutines for the various operations, having every
block at a level send data to every block at the same level, one level
up or down, or (for 2-level algorithms) two levels up or down.  

The number of blocks at each level is not very large, and if a subroutine
determines that the intersection of two blocks is empty (which it does
by examining the bounding rectangles), then it returns immediately.  Thus,
while this algorithm is to be $O(B^2)$ where $B$ is the number of blocks,
it remains $O(N)$ where $N$ is the number of actual points, because
$B$ is much smaller than $N$.

\subsection{block-to-block Subroutines}

The block-to-block subroutines are all designed to take two blocks
(source and destination) and a set of filter coefficients and place
the result of convolving the filter with the source block in the
overlapping points of the destination block.  There is a
block-to-block subroutine for the interpolet transform, its transpose,
its inverse, and its inverse transpose, as well as operator
application routines for the same-level operator, up-one-level
operator, down-one-level operator, up-two-level operator, and the
down-two-level operator.

All of these routines precompute the bounding box for the points in
the destination block which are in the range of influence of the
source block and, for each point in this sub-block, the bounding box
for the points in the source block in the domain of influence of the
destination point.  The result of this precomputation is that the only
data values of the source (destination) which are accessed are the
ones which are read (modified).  This decreases the number of data
accesses in our test problems by a factor of $~7$.

Additionally, blocking the computation generally increases the
locality of access for the data.  More data requests hit the
cache than would occur in a more arbitrarily arranged construction.

\section{Results and Conclusions}

With interpolets, it is possible to carry out $O(N)$ computations in
truncated bases, where $N$ is the number of elements retained in the
truncation, without having to augment the grid of points associated
with the functions maintained in the basis.  Along with allowing one
to compute common linear physical operations, interpolet algorithms
also allow one to transfer between function values and multiscale
expansion coefficients on grids of variable resolution recovering the
same results as one would obtain working with data on a full grid of
arbitrary resolution but without introducing additional grid points
into the calculation.  This allows local nonlinear couplings to be
computed quickly without the introduction of additional sample points
and without the introduction of additional approximations which must
be controlled.

These algorithms have been implemented in Fortran90 and have
subsequently been adopted for use in electronic structure computations
as described in the introduction.  Prior to this, we had been using
very simple, na\"{\i}ve $O(N^2)$ algorithms which implement each
transform and operator as multiplication by the multiscale
representation of the corresponding matrix.  These multiplies check
all points for being within interaction range and then construct the
appropriate matrix element as needed.  This is required in the
na\"{\i}be approach because the variety of inter-scale matrix elements
is too wide to store in table of reasonable size.  This algorithm
ultimately scales quadratically with the number of refinement levels
for our application.  This is because, as described in the
introduction, basis functions are kept in the basis whenever they
contain an atomic nucleus within their support.  All functions in this
subset of significant size of the basis functions associated with each
atomic center therefore touch one another, and the multiscale matrices
contain dense blocks connecting all of these elements of a given
center with one another.  Because the number of functions associated
with a given center grows linearly with the number of refinement
scales $k$, the number of operations required in the na\"{\i}ve
approach of multiplying directly by these dense matrices scales
quadratically with the number of functions in the basis.  For
reference, a typical number of refinement levels in electronic
structure calculations of the lighter elements would be $k=5$, as
employed in the carbon atom \cite{AriasChoLamTeter.proceedings:95} and
the nitrogen molecule \cite{Arias.unpublished:97}.

A comparison with the previous implementation in Fortran90 on the same
processor (Superscalar SPARC Version 9, UltraSPARC) demonstrates the
speed improvements and scaling which can be achieved with the new approach.
The ``time'' axis is the CPU time taken by one application of the
$\nabla^2$ operator.  The ``k'' axis represents the number of levels
of refinement made in the basis and is proportional to the number of
points in $S$.

Figure 9 compares the runtimes of $\nabla^2$ in three dimensions on a
1-level touching good basis with 3rd order interpolets consisting of
concentric cubes of size $15^3$ centered about one atomic nucleus, as
would be appropriate for the calculation of the electronic structure
of a single atom.  Although there is initially a significant $O(N)$
contribution, as a function of the number of refinement levels $k$,
the times for the na\"{\i}ve approach show the constant increments in
slope characteristic of a quadratic function.  The new approach
compares very favorably and is about thirty times faster for typical
values of $k$. (Note the difference in vertical scale between the two
figures.)

Although the comparison in Figure 9 is quite favorable for the new
algorithm, one must bear in mind that given the typical decay in the
interpolet expansion coefficients about an
atom\cite{AriasChoLamTeter.proceedings:95},
\cite{Arias.unpublished:97}, the functions which are appropriate to
maintain in the expansions tend to have the 2-level touching property,
not the 1-level touching property.  Figure 10 compares the runtimes of
$\nabla^2$ in three dimensions on a 2-level touching good basis of
concentric cubes of size $9^3$, where the speed up just as dramatic as
before, now by approximately a factor of 40.

Figure 11 compares the runtimes of $\nabla^2$ in three dimensions on a
2-level touching good basis of two refinement centers, with
refinements now consisting of cubes of size $9^3$ (similar to figure
5).  This situation arises in the the calculation of the electronic
ground state of the nitrogen molecule, $\mbox{N}_2$.  Note that with
the introduction now of two atomic centers the times are again
consistent with the scalings described above: The runs times only
double in the multilevel algorithm but quadruple in the na\"{\i}ve
algorithm.

Having considered the efficiency of the algorithms, we next turn to
the use of these algorithms in the solution of Poisson's equation to
determine electrostatic fields, which was the rate limiting step in
the calculations carried out in \cite{AriasChoLamTeter.proceedings:95}
and \cite{Arias.unpublished:97}.  From those calculations, we were
aware that the combination of conjugate gradients with the simple
preconditioning consisting of just applying the inverse of the
diagonal elements of the Laplacian matrix leads to an algorithm
requiring very few iterations.  It was the application of the operator
within each conjugate gradient iteration which limited the efficiency
of those earlier calculations.

Figures 12-14 illustrate results for varying levels of refinement in
the two different systems.  The first system consists of refinements
centered about a single atomic center within a cubic super cell of
side 15 Bohr radii with periodic boundary conditions.  (One Bohr
radius is approximately 0.529 Angstroms.)  The second system contains
two refinement centers separated at a distance of 2 Bohr radii,
approximately the inter-nuclear separation in the nitrogen molecule.
This latter system resides within a rectangular supercell of
dimensions (15 Bohr)$^2 \times$(17 Bohr).  In both cases, the spacing
of the grid at the coarsest scale is 1 Bohr, and the finest spacing is
$2^{-k}$ Bohr.  At $k=22$, the greatest refinement considered in our
numerical experiments, the finest grid spacing is approximately $0.24
\times 10^{-6}$\AA.  A full grid at this resolution would contain $2.8
\times 10^{23}$ points.  Our truncated basis contains only about
60,000 functions in this case.

Figure 12 compares, as a function of the number of refinement levels
$k$, the condition number of the Laplacian represented in a truncated
interpolet basis (the ``stiffness matrix'' for the basis) and in an
untruncated orthogonal basis at the corresponding resolution.  The
figure also shows the effect on the condition number of the interpolet
stiffness matrix of the simple diagonal preconditioner described
above.  The condition numbers for the truncated interpolet bases were
determined numerically using the operators implemented as described
above.  The curves indicate results for the system with a single
atomic center, and the symbols indicate results for the two atom
system.  Comparing the results for the one and two atom cases suggests
that apart from some transient behavior for small $k$, the
condition number is not sensitive to the number of atoms and depends
primarily on the number of refinement levels $k$.

Although finite basis representations of the Laplacian constructed
from orthogonal functions at a given resolution should all have
similar condition numbers, the fact that the interpolet basis is not
orthogonal allows the condition numbers of multiscale interpolet
operators to be quite different that their single-scale counter parts.
Compared to an orthogonal basis, the condition number in the
interpolet representation is already over two orders of magnitude
superior at the typical $k=5$ levels of refinement.  This comparison
continues to improve with increasing scale.  The orthogonal condition
number scales inversely as the square of the spacing on the grid of
maximum resolution whereas the interpolet condition number scales
inversely with approximately the 5/4 power of the resolution, as
determined from the slope in the figure.  The interpolet basis itself
therefore provides an intrinsic form of preconditioning.  Figure 12
shows that our simple explicit, diagonal preconditioner improves the
scaling of the condition number, which now scales merely as the
inverse of the resolution.  (Note the lower slope of the lower curve.)
At $k=5$ levels of refinement the improvement is only a factor of
three but becomes more significant as the number of refinements
increases.  We extrapolate, based upon the observed scaling behavior,
that the improvement is by a factor of sixty at $k=22$ levels of
refinement, the greatest refinement considered in the examples below.

Figure 13 shows the convergence of the preconditioned conjugate
gradient algorithm in the solution of Poisson's equation.  As a simple
example, we solve for the electrostatic potential which arises from
the two nuclei in a nitrogen molecule.  In this calculation we use
$k=8$ levels of refinement, a somewhat higher level of resolution than
would be employed in calculations of the N$_2$ molecule.  For this
calculation, the charge density of each nucleus is modeled as a three
dimensional Gaussian of root mean square width $\sigma$ along each
direction equal to the spacing on the finest scale.  After an initial
phase of about twenty iterations, the convergence becomes nearly
perfectly exponential.  This procedure reduces the magnitude of the
residual vector by ten orders of magnitude in one hundred iterations.
This is very good performance for a system consisting of 14,000
degrees of freedom with a Laplacian operator with a nominal
single-scale condition number of about 65,000 at this level of
resolution.  The slope of this exponential portion of the convergence
curve corresponds to a reduction in error at each iteration by 25\%.
One would obtain the same error reduction in a simple weighted
iterative Jacobi algorithm (with the inverse of the maximum eigenvalue
as the weight) applied to an operator with condition number $c \approx
4$.  The quantity $c$, the inverse of the fractional improvement in
the magnitude of the residual, we define as the {\em effective
condition number} for the algorithm.

Figure 14 shows this effective condition number $c$ for the conjugate
gradient algorithm with simple diagonal preconditioning as a function
of the number of refinement levels $k$ for the solution of Poisson's
equation for the nuclei in the nitrogen molecule.  In all cases the
extent of the nuclei $\sigma$ is again set to the spacing of the
finest grid.  We note that after about six refinements, the effective
condition number is essentially constant.  The example from Figure 13
is therefore representative of the typical rate of convergence
attained.  These results indicate that, regardless of the
amount of refinement, a constant number of iterations will suffice to
produce a result of a given accuracy, even as the nominal condition
number for an orthogonal stiffness at the corresponding resolution
approaches $1.8 \times 10^{13}$ at $k=22$.  Because the computational work
involved in each iteration is linear in the number of points in the
basis, this approach appears to produce the solution to Poisson's
equation in O(N) time for these multiresolution bases.

\appendix
\newpage
\section{Notation}
\label{notation}

\vskip 4mm
\begin{tabular}{ll}

$C_k$ 		& $2^k \cdot Z^n$ for $k \ge 0$, $Z^n$ for $k < 0$.\\

$D_k$		& $C_{k-1}-C_k$ for $k \ge 0$, $\null$ for $k < 0$.\\

$\level_k$	& $\mbox{min}(k,m)$ where $m$ is the largest integer such that $2^m$
divides all the components of $y$.\\

$\curlyF_\q(S)$	& the space of functions over $S \subset Z^n$.\\

$\curlyI_\q(\phi,S)$	& the space of linear combinations of
$\phi(\frac{x-y}{2^{\level_\q(y)}})$ for $y \in S \subset Z^n$.\\

$\iota^\phi_\q$	& the mapping from $S \rightarrow \curlyI_\q(\phi,S)$ which
takes $y \rightarrow \phi(\frac{x-y}{2^{\level_\q(y)}})$, and linearly
extended to a map from $\curlyF_\q(S) \rightarrow \curlyI_\q(\phi,S)$. \\

$\curlyF_k^S$		& $\curlyF_k(Z^n)/\curlyF_k(Z^n-S).$\\

$\curlyI_k^S(\phi)$	& $\curlyI_k(\phi,Z^n)/\curlyI_k(\phi,Z^n-S).$\\
$\tilde{v}$	& the zero-lift representative of $v \in \curlyF^S_k$, i.e. 

$\tilde{v} \in \curlyF_k$, such that $\tilde{v}(y) =
v(y), y\in S$ and $\tilde{v}(y) = 0, y\notin S$. \\

$V^*$		& the dual space of the vector space $V$.\\

$J_{k_1, k_2}$	& the map, $\curlyF_{k_1} \rightarrow \curlyF_{k_2},$
given by $J_{k_1,k_2} = \iota_{k_2}^{-1} \circ \iota_{k_1}.$ \\

$J_k$		& short for $J_{0,k}.$ \\

$J_{-k}$	& short for $J_{k,0}.$ \\

$\left< f | {\cal O} | g\right>$	& the matrix element $\int f(x) {\cal O} g(x) d^nx.$ \\

\end{tabular}

\newpage
\section{Table of Contents}
\label{toc}

\tableofcontents

\newpage
\bibliography{ref}
\bibliographystyle{plain}

\newpage

\noindent
{\large Figures: (available from
http://laisla.mit.edu/muchomas/Papers/nonstand-figs.ps or in the file
``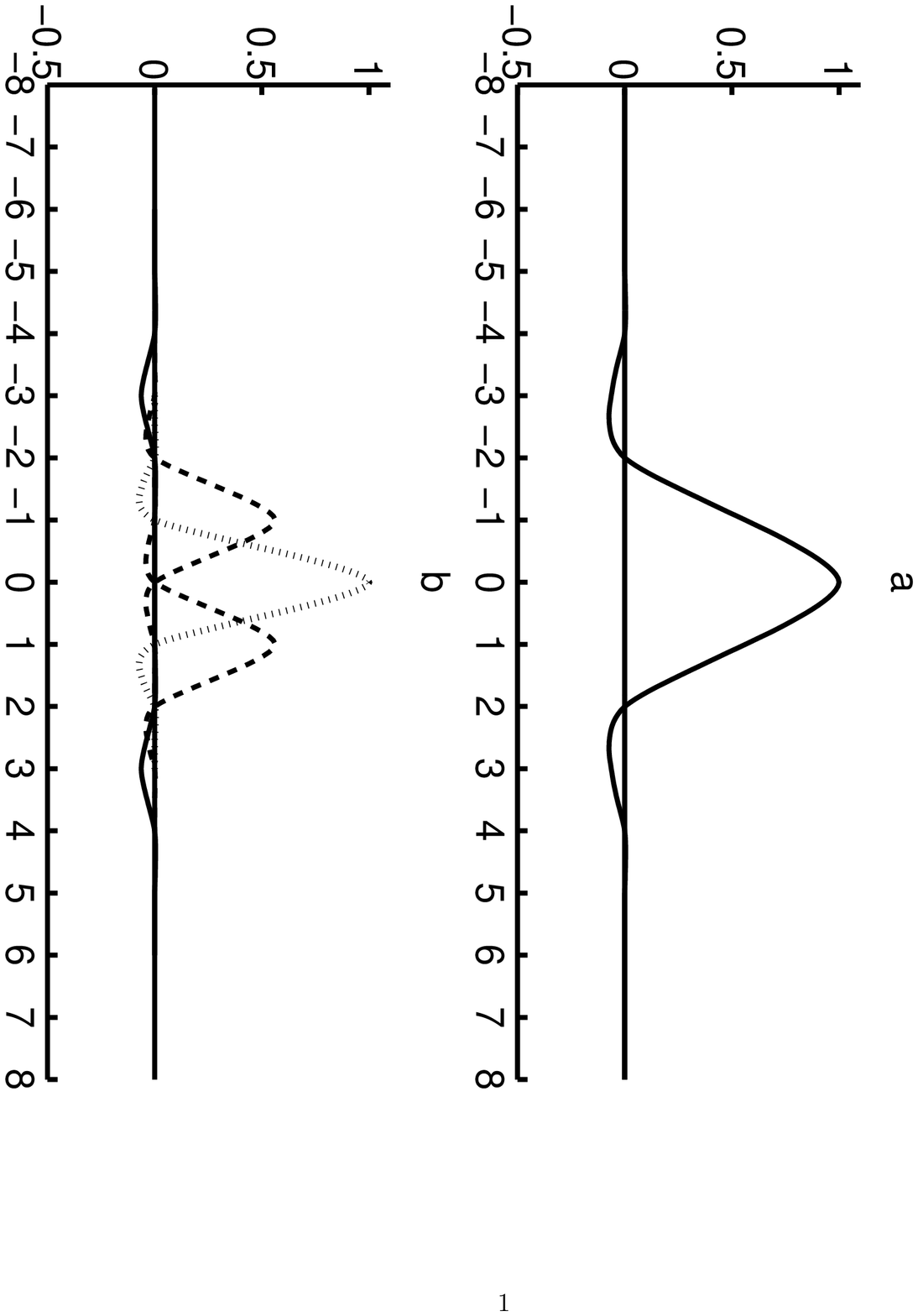'' in this posting)}

Figure 1: (a) shows $\phi(x/2)=\frac{-1}{16}\phi(x-3)+
\frac{9}{16}\phi(x-1)+\phi(x) + \frac{9}{16} \phi(x+1) + 
\frac{-1}{16} \phi(x+3)$.  (b) shows the functions
$\frac{-1}{16}\phi(x-3),\frac{9}{16}\phi(x-1), \phi(x) , 
\frac{9}{16} \phi(x+1) , \frac{-1}{16} \phi(x+3)$.  

Figure 2: (a) the smooth function $f(x) = 
(\frac{x}{2}+3)^3e^{-(\frac{x}{4})^4}$. (b) the approximation 
to $f(x)$ in $\curlyI_0(Z)$. (c) component of the approximation 
in $\curlyI_1(C_1)$.  (d) component of the approximation in 
$\curlyI_1(D_1)$.

Figure 3: Approximation of the function $f(x) = e^{-|x|}$
with interpolets centered at the tick marks: 
(a) a graph of $e^{-|x|}$.
(b) the cardinal approximation constructed from interpolets centered 
at the knots of a uniform grid.
(c) and (d) cardinal approximations from local refinements of the uniform grid.
The additional levels of refinement near the singularity increase the 
accuracy.

Figure 4: (left) a graph of the ${\cal L}^2$ relative error for $k$ levels of local dyadic refinement.
(right)  a graph of the ${\cal L}^\infty$ relative error for $k$ levels of local dyadic refinement.

Figure 5: An example of a truncation one might wish
to use for a diatomic molecule (two atomic cores).  In black are shown
those points in $S \subset Z^n$ whose residual values might
be significant.

Figure 6: A visual summary of the 1-level touching condition.
Solid plots represent functions centered at points in $S$.
Dotted plots represent functions centered at points in $Z^n-S$.  Tick
marks delimit the overlap of the two functions.

Figure 7: Our example of the 1-level touching good basis 
in one dimension.  Note that the two functions plotted do not touch.

Figure 8: A generic example of a truncation 
which meets our definitions of good and 1-level touching.

Figure 9: The previously used
implementation is on the left,
and the implementation employing a 1-level touching algorithm
is on the right.  (Note the difference in scale on the vertical axes.)

Figure 10: The previously used
implementation is on the left,
and the implementation employing a 2-level touching algorithm
is on the right.  (Note the difference in scale on the vertical axes.)

Figure 11: The previously used
implementation is on the left,
and the implementation employing a multilevel algorithm
is on the right.  (Note the difference in scale on the vertical axes.)

Figure 12: The condition number of the Laplacian operator represented
in truncated multiresolution interpolet basis as a function of the
number of refinement levels $k$ with and without simple diagonal
preconditioning and compared with the condition number in an
orthogonal basis with the same resolution.  Lines indicate results for
bases with a single atomic center of refinement and points represent
results for two atomic centers corresponding to the nitrogen molecule.

Figure 13: Convergence of the solution to Poisson's equation for the
nuclear potential in a nitrogen molecule in an interpolet basis with
$k=8$ levels of refinement about each nucleus.

Figure 14: Effective condition number of Poisson's equation for the
nuclear potential in a nitrogen molecule with simple diagonal
preconditioning as a function of $k$ the number of levels of
refinement about each nucleus.

\end{document}